\documentclass{ieeeaccess}
\usepackage{cite}
\usepackage{physics}   
\usepackage{amsmath,amssymb,amsfonts}
\usepackage{algorithmic}
\usepackage{graphicx}
\usepackage{textcomp}
\usepackage{braket}
\usepackage{verbatim}
\def\BibTeX{{\rm B\kern-.05em{\sc i\kern-.025em b}\kern-.08em
    T\kern-.1667em\lower.7ex\hbox{E}\kern-.125emX}}
\begin{document}
\history{Date of publication xxxx 00, 0000, date of current version xxxx 00, 0000.}
\doi{10.1109/ACCESS.2022.0122113}

\title{A Step-by-Step HHL Algorithm Walkthrough to Enhance Understanding of Critical Quantum Computing Concepts}
\author{\uppercase{Anika Zaman}\authorrefmark{1},
\uppercase{Hector Jose Morrell\authorrefmark{2}, Hiu Yung Wong}.\authorrefmark{3},
\IEEEmembership{Senior Member, IEEE}}
\address[1]{Department of Electrical Engineering, San Jose State University, San Jose, CA 95192 USA (email: anika.zaman@sjsu.edu)}
\address[2]{Department of Electrical Engineering, San Jose State University, San Jose, CA 95192 USA (email: hector.morrell@sjsu.edu)}
\address[3]{Department of Electrical Engineering, San Jose State University, San Jose, CA 95192 USA (email: hiuyung.wong@sjsu.edu)}
\tfootnote{Part of this work was supported by the National Science
Foundation under Grant 2046220.}

\markboth
{Author \headeretal: }
{Author \headeretal: }

\corresp{Corresponding author: Hiu Yung Wong (email: hiuyung.wong@sjsu.edu). A. Zaman and H. J. Morrell have equal contributions.}

\begin{abstract}
After learning basic quantum computing concepts, it is desirable to reinforce the learning using an important and relatively complex algorithm through which students can observe and appreciate how qubits evolve and interact with each other. Harrow-Hassidim-Lloyd (HHL) quantum algorithm, which can solve linear system problems with exponential speed-up over the classical method and is the basis of many important quantum computing algorithms, is used to serve this purpose. The HHL algorithm is explained analytically followed by a 4-qubit numerical example in bra-ket notation. Matlab code corresponding to the numerical example is available for students to gain a deeper understanding of the HHL algorithm from a pure matrix point of view. A quantum circuit programmed using qiskit is also provided for real hardware execution in IBM quantum computers. After going through the material, students are expected to have a better appreciation of the concepts such as basis transformation, bra-ket and matrix representations, superposition, entanglement, controlled operations, measurement, quantum Fourier transformation, quantum phase estimation, and quantum programming. To help readers review these basic concepts, brief explanations augmented by the HHL numerical examples in the main text are provided in the Appendix.

\end{abstract}

\begin{keywords} Harrow-Hassidim-Lloyd (HHL) quantum algorithm, 
Quantum Fourier transform (QFT), inverse quantum Fourier transform (IQFT), quantum phase estimation(QPE),linear system problem (LSP), Quantum Education
\end{keywords}

\titlepgskip=-15pt

\maketitle

\section{Introduction}
\label{sec:introduction}
\PARstart{Q}{uantum} Computing is promising in solving challenging engineering \cite{Shor1994}, biomedical \cite{Outeiral} and finance \cite{Egger2020} problems. It has a tremendous advancement in the last two decades and, recently, quantum breakthrough has been demonstrated using a 53-qubit system \cite{ Arute2019}.However, according to the paper \cite{ Madsen2022}, due to less efficient hardware implementation till date, the goal to reach superconducting quantum supremacy is yet to achieve. Therefore, the training of a quantum technology workforce is an imminent goal for many countries (e.g. \cite{NSTC}) to support this fast-growing industry. 
However, quantum technology is based on concepts very different from our daily and classical experiences. In the early stage of learning quantum computing, although linking to daily and classical experience may enhance the understanding of certain quantum concepts and such an approach should not be de-emphasized, we believe a fast and robust way of training a quantum workforce is to train the students to be able to emulate a quantum processor and trace the evolution of the qubits. This is particularly useful in learning quantum algorithms without a quantum mechanics background. Such an approach obviates the students from cognitive conflicts, which can be resolved later, if possible, after they understand how quantum computing works. This also embraces the ``Shut up and calculate!” approach proposed by Mermin on how to deal with the uncomfortable feeling toward quantum mechanics interpretation \cite{ Mermin2004}.

Besides analytical equations, matrix representation and computer simulations are important tools to enhance the understanding of qubit evolution. However, available examples that include computer simulations are usually of simple algorithms and, very often, without matrix representation. There is a lack of examples of important and relatively complex algorithms which combine some of the most important quantum computing concepts and basic algorithms. Such examples are desirable to allow students to appreciate the roles and the interplay of various basic concepts in a more realistic quantum algorithm. Harrow-Hassidim-Lloyd (HHL) quantum algorithm \cite{HHL}\cite{Cao} which can be used to solve linear system problems (LSP) and can provide exponential speedup over the classical conjugate gradient method \cite{Chandra1975} is chosen for this purpose. HHL is the basic of many more advanced algorithms and is important in various applications such as machine learning \cite{Dmitry} and modeling of quantum systems \cite{Outeiral}\cite{survey}. 
HHL solves system of linear equation which is a discretization of  \cite{Wang}\cite{sispad}. 
In this paper, we detail the qubit evolution in 
Harrow-Hassidim-Lloyd (HHL) quantum algorithm analytically with a 4-qubit circuit as a numerical example. Although HHL examples are available elsewhere (e.g. \cite{Schleich}\cite{HHL_ex}), this paper has certain characteristics which are not all found in those examples. Firstly, the HHL algorithm is discussed analytically step-by-step and is self-contained. Secondly, a numerical example is given in bra-ket notation mirroring the analytical equations. Thirdly, a Matlab code corresponding exactly to the numerical example is available to enhance the understanding from a matrix point of view. The Matlab code allows the students to trace how the wavefunction evolves instead of just seeing the magnitudes of the coefficients as in IBM-Q. Fourthly, a qiskit code written in python \cite{ qiskit} corresponding to the numerical example is available and can be run in IBM simulation and hardware machines \cite{ibm}. Finally, in the example, all the 4 qubits are traced throughout the process without simplification.

The readers are assumed to have the following background concepts which are further enhanced through the step-by-step walkthrough of the HHL algorithm: basis transformation, bra-ket and matrix representations, superposition, entanglement, encoding, controlled operations, measurement, quantum Fourier transformation, and quantum programming. To make this paper self-contained and to help the readers better appreciate the roles of these basic concepts in the HHL, an Appendix is devoted to briefly explaining these concepts using the examples from the main text. A more detailed explanation of these concepts using a similar approach as in this paper can be found in \cite{Wong2022}.

\subsection{How to use this paper}

For readers who have a fresh memory of the basic concepts, they can start reading from Section~\ref{HHL_Algorithm}, in which the HHL algorithm is discussed step-by-step analytically followed by a numerical example in Section~\ref{numerical_example}. The basic concepts mentioned in the Appendix are referred to in the main text and readers are encouraged to review them when needed.

For readers who need reviews on the basic concepts first, they are encouraged to go over the Appendix first before reading the main text.  

For readers who have devoted substantial time to learning HHL elsewhere but just need a numerical example to reinforce the understanding, they might start with the numerical example in Section~\ref{numerical_example}.

Equations in the Appendix begin with '\ref{appendix}'. If the equations are examples from the main text, the same equation number is used in the Appendix.

\section{HHL Algorithm} \label{HHL_Algorithm}

\subsection{Definitions and Overview}

We will first give an overview of the problem and the HHL algorithm. Details will be discussed in the following subsections with reference to the Appendix for reviewing basic concepts.
A linear system problem (LSP) can be represented as the following 
\begin{equation}
\label{tosolve}
A\vec{x}=\vec{b}
\end{equation}
where $A$ is a $N_b\times N_b$ Hermitian matrix and $\vec{x}$ and $\vec{b}$ are $N_b$-dimensional vectors. For simplicity, it is assumed $N_b=2^{n_b}$, where $n_b$ is the number of qubits in the quantum circuit, and $N_b$ is the total combinations due number of qubits,$n_b$. In matrix representation, qubits are represented by their total combinations($N_b \times N_b$), or we can say for $N_b$ number of unknowns, we need $n_b$ qubits to solve unknowns. Dummy equations can be added otherwise to convert the system satisfy this assumption. $A$ and $\vec{b}$ are known and $\vec{x}$ is the unknown to be solved, i.e.

\begin{equation}
\label{linearsol}
\vec{x}=A^{-1}\vec{b}
\end{equation}

\begin{figure*}[h!]
\centering
\includegraphics[width=40em]{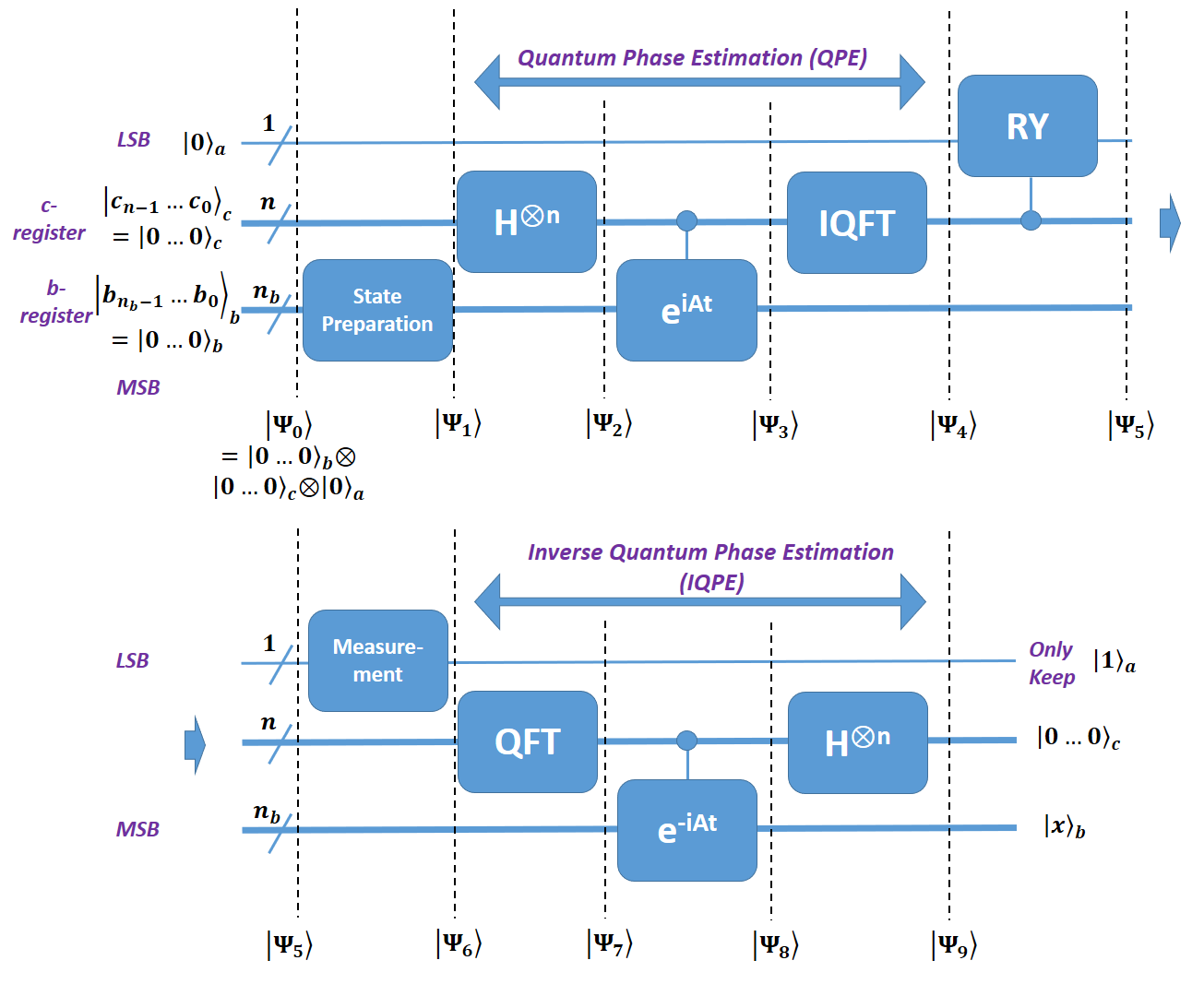}
\caption{Schematic of the $HHL$ quantum circuit flowing from left to right. The circuit is decomposed into top and bottom portions for clarity. Note that the lowest qubit in the diagram is the most significant bit ($MSB$) while the top one is the least significant bit ($LSB$).}
\label{Schematic}
\end{figure*}
As an example, $A = \begin{pmatrix} 1 & -\frac{1}{3} \\ -\frac{1}{3} & 1 \\ \end{pmatrix}$, $\vec{b} = \begin{pmatrix} 0 \\ 1 \\ \end{pmatrix}$, and $\vec{x} = \begin{pmatrix} \frac{3}{8} \\ \frac{9}{8} \\ \end{pmatrix}$ with $n_b=1$ and $N_b=2^1=2$. Readers may refer to Appendix~\ref{A_Gaussian} and Appendix~\ref{A_conjugate} to review how LSP is solved classically using Gaussian Elimination and Conjugate Gradient Method, respectively.

$A$ is assumed to be Hermitian (See Appendix~\ref{A_Hermitian}). If it is not Hermitian, the $A$ can be converted to  $\begin{pmatrix} 0 & A \\ A^\dagger & 0 \\\end{pmatrix}$, which is Hermitian. Readers may refer to \cite{Dervovic2018} for the more advanced treatment when $A$ is not Hermitian.

Figure~\ref{Schematic} shows the schematic of the $HHL$ algorithm and the corresponding circuit to solve  LSP. In the HHL quantum algorithm, the $N_b$ components of $\vec{b}$ and $\vec{x}$ are encoded as the amplitudes/coefficients (\emph{amplitude encoding}) of basis states of the $n_b$-qubits, $\ket{\;}_{b}$, which form a $\mathbb{C}^{N_b}$ Hilbert space. These $n_b$ qubits are called b-register. Qubit $n_b$ is chosen to be large enough to encode $\vec{b}$, i.e. $2^{n_b}$ needs to be the same as the length of the vectors $\vec{b}$ and $\vec{x}$. The matrix $A$ is simulated through \emph{Hamiltonian encoding} by encoding it as the Hamiltonian of a unitary gate. Appendix~\ref{A_Encoding} reviews examples of various encoding schemes. 

The HHL algorithm has 5 main components, namely state preparation, quantum phase estimation (QPE), ancilla bit rotation, inverse quantum phase estimation (IQPE), and measurement. In this paper, the little-endian convention is used. In a little-endian convention, the rightmost (ending) qubit represents the least significant bit (LSB). For example, in a 4-qubit system, $\ket{0001}$ in binary is $\ket{1}$ in decimal because the $1$ in the basis state $\ket{0001}$ is the LSB, representing $2^0$ instead of $2^3$ (if it were the most significant bit, MSB). Moreover, in the circuit diagrams, the lowest qubit represents the MSB and the topmost qubit represents the LSB, which is a convention used in qiskit \cite{qiskit} and the IBM-Q platform \cite{ibm}

As shown in Figure~\ref{Schematic}, besides the b-register, which belongs to the more significant bits, there are two more sets of inputs to the algorithm. The first set is sometimes called the c-register because it is related to the time (clock) in the controlled rotation in the QPE part. Therefore, they are also called the clock qubits.The c-register stores the values of the phase of the eigenvalues of the $A$ matrix after the QPE. There are $n$ qubits in the c-register. Since \emph{basis encoding} is used (i.e. the phase value is encoded as the basis number (See Appendix~\ref{A_Encoding}), the value of $n$ determines how accurately the phase can be stored. A larger $n$ results in higher accuracy when the encoding is not exact. We set $N=2^n$.

The last set of qubits is the ancilla qubit $\ket{\;}_{a}$ which is the LSB. The ancilla qubit, as its name implies, is important to help achieve the goal although it will be discarded at the end, as will be detailed later.

The matrix $A$, which is a Hamiltonian, may be written as a linear combination of the outer products of its eigenvectors, $\ket{u_i}\bra{u_i}$ weighted by its eigenvalues, $\lambda_i$,in Eq.(\ref{eq:3}), (See Appendix \ref{A_Eigenvalues}). 

\begin{equation}
\label{eq:3}
A=\sum_{i=0}^{2^{n_b}-1} \lambda_i \ket{u_i} \bra{u_i}
\end{equation}

Since A is diagonal in its eigenvector basis, its inverse is simply, $A^{-1}=\sum_{i=0}^{2^{n_b}-1} \lambda_i^{-1} \ket{u_i} \bra{u_i}$. $\vec{b}$ can be also expressed in the basis formed by the eigenvectors of A, such that 
\begin{equation}
\label{b_in_A_basis}
\ket{b}=\sum_{j=0}^{2^{n_b}-1}b_j \ket{u_j} 
\end{equation}
Therefore, Eq.~(\ref{linearsol}) can be encoded as,

\begin{equation}
\label{xsolution}
\ket{x}=A^{-1}\ket{b}=\sum_{i=0}^{2^{n_b}-1} \lambda_i^{-1}b_i \ket{u_i}
\end{equation}

\noindent by using the fact that $\braket{u_i|u_j}=\delta_{ij}$. The goal of the HHL algorithm is to find the solution in this form and $\ket{x}$ is stored in the b-register.
Storing the right hand side of (\ref{xsolution}) in the b-register is equivalent to storing $\ket{x}$ in the b-register. But the solution is encoded as the amplitudes of the basis vectors $\ket{0}/\ket{1}$ (the measurement basis). Therefore, the solutions are not $\lambda_i^{-1}b_i$, which are the amplitudes of the eigenvector basis vectors. However, one will naturally get the correct amplitudes if it is measured in the $\ket{0}/\ket{1}$ basis and this is only possible if the qubits are not entangled with other qubits. This can be checked mathematically replacing $\ket{u_i}$ by $\ket{0}/\ket{1}$ based on their relationship. If $\ket{u_i}$ are entangled with other qubits, one cannot obtain the desired solution. This will be clear after the ancilla bit rotation to be detailed later.

The equation needs also to be prepared so that the eigenvectors, $\ket{u_i}$, and $\ket{b}$ are normalized so they can be properly represented as a unit vector in quantum computing. Therefore, (\ref{b_in_A_basis}) and Eq.~(\ref{xsolution}) require
\begin{align}
\sum_{j=0}^{2^{n_b}-1}\lvert b_j\rvert^2=1 \nonumber\\
\label{eq6}
\end{align}
\begin{align}
\sum_{i=0}^{2^{n_b}-1} \lvert\lambda_i^{-1}b_i\rvert^2=1
\label{eq7}
\end{align}

\subsection{State Preparation} 

There are total $n_b+n+1$ qubits and they are initialized as

\begin{equation}
\ket{\Psi_0} = \ket {0\cdots0}_{b}\ket {0\cdots0}_{c}\ket {0}_a=\ket {0}^{\otimes n_b}\ket {0}^{\otimes n}\ket {0}
\end{equation}

In the state preparation, $\ket {0\cdots0}_{b}$ in the b-register needs to be rotated to have the amplitudes correspond to the coefficients of $\vec{b}$. That is 

\begin{equation}
\vec{b}=
\begin{pmatrix}
\beta_0 \\
\beta_1 \\
\vdots \\
\beta_{N_b-1} \\
\end{pmatrix}
\Leftrightarrow
\beta_0\ket{0}+\beta_1\ket{1}+\cdots+\beta_{N_b-1}\ket{N_b-1}=\ket{b}
\label{b-representation}
\end{equation}

The vector $\vec{b}$ is represented in a column form on the left with coefficients $\beta's$. This is also a valid representation of $\ket{b}$. On the right, the corresponding basis of the Hilbert space formed by the $n_b$ qubits is written explicitly. Therefore,

\begin{equation}
\ket{\Psi_1}= \ket {b}_b\ket {0\cdots0}_{c}\ket {0}_a
\end{equation}

From now on, some of the subscripts of the kets will be omitted when there is no ambiguity. Since the state preparation depends on the actual value of $\vec{b}$, it will be discussed in more detail in the numerical example.

\subsection{Quantum Phase Estimation} 

Quantum phase estimation (QPE) is also an eigenvalue estimation algorithm. QPE has three components, namely the superposition of the clock qubits through Hadamard gates, controlled rotation, and inverse quantum Fourier transform (IQFT). The goal of QPE is to estimate the phase of the eigenvalues of the unitary rotation matrix, $U=e^{iAt}$, in the controlled gate, $C-U$, (Fig.~\ref{Schematic}) used in the QPE. Again, this gate encodes the matrix $A$ as its Hamiltonian. It is also instructive to note that the eigenvalues of $U$ must be roots of unity (i.e. in the form of $e^{i\theta}$) as $U$ is unitary. Therefore, the phase of the eigenvalue of the gate is proportional to the eigenvalue of $A$. As a result, by using QPE in the HHL algorithm, it is expected the eigenvalues of $A$ will be encoded in the c-register after the QPE, i.e. at $\ket{\Psi_4}$. As it will be clear later, the eigenvalues are only encoded through basis encoding. The c-register does not store the exact eigenvalues.

Here we assume the readers are already familiar with IQFT and it will not be explained in detail. Readers may review the basic concepts in Appendices~\ref{A_QFT} and \ref{A_implement_QFT}.

In the first step of QPE, Hadamard gates are applied to the clock qubits to create a \emph{superposition} of the clock qubits,
\begin{align}\label{eq:9-10}
\ket{\Psi_2}
& = & I^{\otimes n_b}\otimes H^{\otimes n} \otimes I\ket{\Psi_1}\\
& = & \ket{b}\frac{1}{2^{\frac{n}{2}}}(\ket{0}+\ket{1})^{\otimes n}\ket{0}
\end{align}

\begin{figure}[h!]
\centering
\includegraphics[width=3.6in]{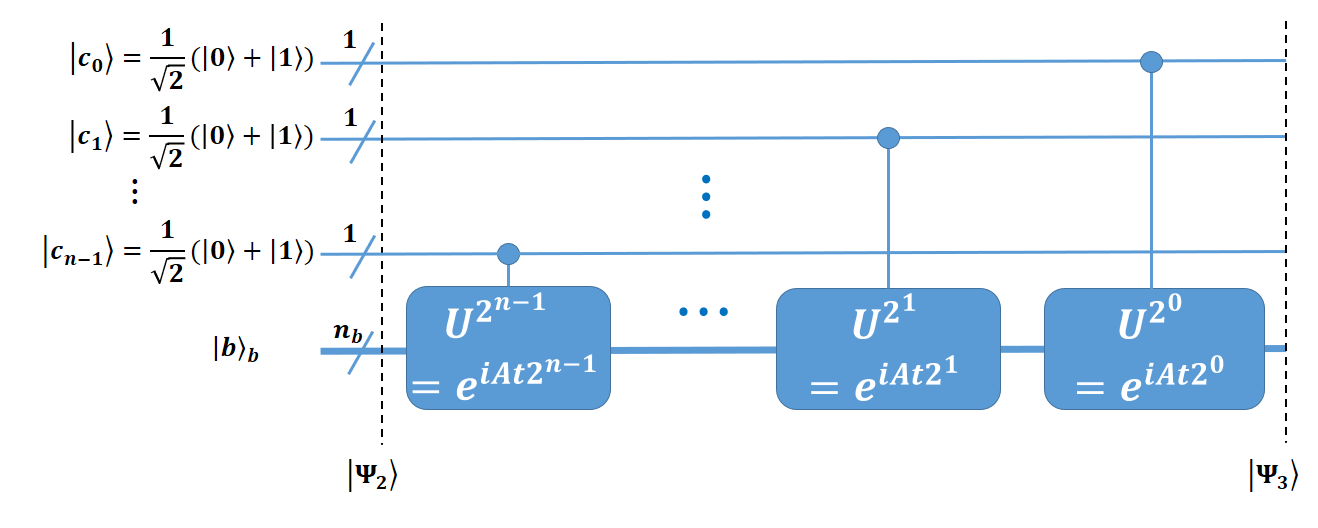}
\caption{The controlled-rotation part of QPE. $U$ is replaced by $e^{iAt}$ in the HHL algorithm.}
\label{QPE-controlled}
\end{figure}

In the controlled rotation part, controlled gates are applied to $\ket{b}$ with the clock qubits as the controlling qubits (Figure~\ref{QPE-controlled}). The number of the qubit, $n$, of the c-register determines the number of the controlled gates. The gates are in the form of $U^{2^r}$, where $r$ is the index of the clock qubit. Also, $U=e^{iAt}$. For the most significant bit in the c-register, $\ket{c_{n-1}}$, it controls the gate $U^{2^{n-1}}$ on the b-register while the least significant one, $\ket{c_{0}}$, controls the gate $U^{2^0}=U$ on the b-register.

To understand how it works, we begin by assuming that $\ket{b}$ is an eigenvector of $U$ with eigenvalue $e^{2\pi i \phi}$. The eigenvalue is written in this form so that the phase, $\phi$, will be encoded as the basis state in the c-register (See Eq.~(\ref{phi4_approx})).  Therefore, based on the definition of eigenvalues and eigenvectors (See Section~\ref{A_Eigenvalues}), 
\begin{equation}
\label{Ub}
U\ket{b}=
e^{2\pi i \phi}\ket{b}
\end{equation}

When the control clock qubit is $\ket{0}$, $\ket {b}$ will not be affected. If the clock bit is $\ket{1}$, $U$ will be applied to $\ket {b}$. This is equivalent to multiplying $e^{2\pi i \phi 2^{j}}$ in front of the $\ket{1}$ of the $j${\rm th} clock qubit, $\ket{c_j}$, as one can assign the prefactor to the controlling qubit. Therefore, after the controlled-$U$ operation, we have 

\begin{align}
\label{eq:14}
\ket{\Psi_3}
& =  \ket {b}\otimes\big(\frac{1}{2^{\frac{n}{2}}}(\ket{0}+e^{2\pi i \phi 2^{n-1}}\ket{1})\otimes (\ket{0}+\nonumber\\
& e^{2\pi i \phi 2^{n-2}}\ket{1})\otimes  
\cdots \ \otimes (\ket{0}+e^{2\pi i \phi
2^{0}}\ket{1})\big)\otimes\ket{0}_a \nonumber \\
& =  \ket {b}\frac{1}{2^{\frac{n}{2}}} \sum_{k=0}^{2^{n}-1} e^{2\pi i \phi k}\ket{k}\ket{0}_a
\end{align}

In the IQFT part,(\ref{eq:13}), only the clock qubits are affected. Note that in certain literature, this is called Quantum Fourier Transform (QFT) (Appendix~\ref{A_QFT}).
\begin{align}
\label{eq:13}
\ket{\Psi_4}
& =  \ket {b} \textrm{IQFT}(\frac{1}{2^{\frac{n}{2}}} \sum_{k=0}^{2^{n}-1} e^{2\pi i \phi k}\ket{k})\ket{0}_a\nonumber\\
& =  \ket {b} \frac{1}{2^{\frac{n}{2}}} \sum_{k=0}^{2^{n}-1} e^{2\pi i \phi k}(\textrm{IQFT}\ket{k})\ket{0}_a\nonumber\\
& =  \ket {b} \frac{1}{2^{n}}\sum_{k=0}^{2^{n}-1}e^{2\pi i \phi k}(\sum_{y=0}^{2^{n}-1} e^{-2\pi i y k/N} \ket{y})\ket{0}_a\nonumber\\
& = \frac{1}{2^{n}}\ket {b}\sum_{y=0}^{2^{n}-1} \sum_{k=0}^{2^{n}-1} e^{2\pi i k(\phi -y/N)}\ket{y}\ket{0}_a
\end{align}

Due to \emph{interference}, only $\ket{y}$ satisfying the condition $\phi -y/N = 0$ will have a finite amplitude of $\sum_{k=0}^{2^{n}-1} e^{0}=2^n$. Otherwise, the amplitude is $\sum_{k=0}^{2^{n}-1} e^{2\pi i k(\phi -y/N)}=0$ due to \emph{destructive interference}. By ignoring the states with zero amplitude, we may rewrite $\ket{\Psi_4}$ as
\begin{align}
\label{phi4_approx}
\ket{\Psi_4}
& =  \frac{1}{2^{n}}\ket {b}\sum_{k=0}^{2^{n}-1} e^{2\pi i k\cdot 0}\ket{N\phi}\ket{0}_a\nonumber\\
&=\ket {b}\ket{N\phi}\ket{0}_a
\end{align}

Therefore, in QPE, the clock qubits are used to represent the phase information of $U$, which is $\phi$, and the accuracy depends on the number of qubits, $n$.

Since in \emph{Hamiltonian encoding}, $U$ is related to $A$ through

\begin{equation}
U=e^{iAt}
\label{eq:15}
\end{equation}
where $t$ is the evolution time for that Hamiltonian. $U$ is also diagonal in $A's$ eigenvector, $\ket{u_i}$, basis. If $\ket{b}=\ket{u_j}$, 
\begin{eqnarray}
U\ket{b}
& = &e^{i\lambda_j t}\ket{u_j}
\end{eqnarray}

By equating $i\lambda_j t$ to $2\pi i \phi$ in Eq.~(\ref{Ub}), we get $\phi=\lambda_j t/2\pi$ and Eq.~(\ref{phi4_approx}) becomes

\begin{equation}
\ket{\Psi_4}=\ket {u_j}\ket{N\lambda_j t/2\pi }\ket{0}_a
\end{equation}

Thus the eigenvalues of $A$ have been encodeded in the clock qubits (\emph{basis encoding}). However, in general, given in (\ref{b_in_A_basis}), by \emph{superposition},
\begin{equation}
\ket{\Psi_4}=\sum_{j=0}^{2^{n_b}-1}b_j\ket {u_j}\ket{N\lambda_j t/2\pi }\ket{0}_a
\end{equation}

The $\lambda_j$ are usually not integers. We will choose $t$ so that $\tilde{\lambda_j}=N\lambda_j t/2\pi$ are integers. Therefore, $\tilde{\lambda_j}$ are usually scaled version of $\lambda_j$. 

$\Psi_4$ can be rewritten as

\begin{equation}
\label{finalPsi4}
\ket{\Psi_4}=\sum_{j=0}^{2^{n_b}-1}b_j\ket {u_j}\ket{\tilde{\lambda_j}}\ket{0}_a
\end{equation}

\subsection{Controlled Rotation and Measurement of the Ancilla Qubit}

The next step is to rotate the ancilla qubit, $\ket{0}_a$, based on the encoded eigenvalues in the c-register, such that,

\begin{equation}
\label{ancilla_R}
\ket{\Psi_5}=\sum_{j=0}^{2^{n_b}-1}b_j\ket {u_j}\ket{\tilde{\lambda_j}}(\sqrt{1-\frac{C^2}{\tilde{\lambda_j}^2}}\ket{0}_a+\frac{C}{\tilde{\lambda_j}}\ket{1}_a)
\end{equation}
where $C$ is a constant. The goal is to show why this is useful.

When the ancilla qubit is measured, the ancilla qubit \emph{wavefunction} will \emph{collapse} to either $\ket{0}$ or $\ket{1}$. If it is $\ket{0}$, the result will be discarded and the computation will be repeated until the measurement is $\ket{1}$. Therefore, the final wavefunction of interest is

\begin{equation}
\ket{\Psi_6}=\frac{1}{\sqrt{\sum_{j=0}^{2^{n_b}-1}\lvert\frac{b_j C}{\tilde{\lambda_j}}\rvert^2}}\sum_{j=0}^{2^{n_b}-1}b_j\ket {u_j}\ket{\tilde{\lambda_j}}\frac{C}{\tilde{\lambda_j}}\ket{1}_a
\end{equation}
where the prefactor is due to normalization after measurement. Since $\lvert\frac{C}{\tilde{\lambda_j}}\rvert^2$ is the probabily of obtaining $\ket{1}$ when the ancilla bit is measured, $C$ should be chosen to be as large as possible. Compared to (\ref{xsolution}), the result resembles the answer $\ket{x}$ that we are looking for. However, we can only obtain the correct result if the b-register is measured in the eigenvector basis (i.e. $\ket {u_j}$ instead of $\ket{0}/\ket{1}$). However, the b-register is \emph{entangled} with the clock qubits, $\ket{\tilde{\lambda_j}}$. This means that we cannot factorize the result into a tensor product of the c-register and b-register (See the discussion after (\ref{xsolution}) and Appendix~\ref{entanglement}). As a result, we cannot convert the b-register into the $\ket{0}/\ket{1}$ measurement basis with the desired amplitudes. We will need to uncompute the state so that it gives the right results in the $\ket{0}/\ket{1}$ measurement during which the b-register and c-register will be \emph{unentangled}.

The measurement of the ancilla qubit can be and is usually performed after uncomputation. However, since the ancilla bit is not involved in any operations after the controlled rotation, measuring the ancilla bit before the uncomputation gives the same result. For simplicity in the derivation, it is thus performed before the uncomputation.

\subsection{Uncomputation - Inverse QPE}

The uncomputation is done by using inverse QPE. Firstly, QFT is applied to the clock qubits as, 
\begin{align}
\ket{\Psi_7}
& =  \frac{1}{\sqrt{\sum_{j=0}^{2^{n_b}-1}\lvert\frac{b_j C}{\tilde{\lambda_j}}\rvert^2}}\sum_{j=0}^{2^{n_b}-1}\frac{b_j C}{\tilde{\lambda_j}}\ket {u_j}QFT\ket{\tilde{\lambda_j}}\ket{1}_a\nonumber\\
& =  \frac{1}{\sqrt{\sum_{j=0}^{2^{n_b}-1}\lvert\frac{b_j C}{\tilde{\lambda_j}}\rvert^2}}\sum_{j=0}^{2^{n_b}-1}\frac{b_j C}{\tilde{\lambda_j}}\ket {u_j}\nonumber \\
& \cdot(\frac{1}{2^{n/2}}\sum_{y=0}^{2^{n}-1} e^{2\pi i y\tilde{\lambda_j}/N} \ket{y})\ket{1}_a
\end{align}

Then inverse controlled-rotations of the b-register by the clock qubits are applied with $U^{-1}=e^{-iAt}$. Similar to the forward process, when the controlling $r$\textrm{th} clock qubit is $\ket{0}$, $\ket {u_j}$ will not be affected. If the $r$\textrm{th} clock qubit is $\ket{1}$, $(U^{-1})^{2^r}$ will be applied to $\ket {u_j}$. This is equivalent to multiplying $e^{-i\lambda_j t y}$ if the c-register is $\ket{y}$. This is due to the similar argument in Eq.~(\ref{eq:14}) and the fact that $2\pi i \phi=i\lambda_j t$. Therefore,

\begin{align}
\ket{\Psi_8}
& = & \frac{1}{2^{n/2}\sqrt{\sum_{j=0}^{2^{n_b}-1}\lvert\frac{b_j C}{\tilde{\lambda_j}}\rvert^2}}\sum_{j=0}^{2^{n_b}-1}\frac{b_j C}{\tilde{\lambda_j}}\ket {u_j}\nonumber\\
&&\cdot(\sum_{y=0}^{2^{n}-1} e^{-i\lambda_j ty}e^{2\pi i y\tilde{\lambda_j}/N} \ket{y})\ket{1}_a
\end{align}

Since we earlier chose to set $\tilde{\lambda_j}=N\lambda_j t/2\pi$, therefore, the two exponential terms cancel each other and 

\begin{align}
\ket{\Psi_8}
& = & \frac{1}{2^{n/2}\sqrt{\sum_{j=0}^{2^{n_b}-1}\lvert\frac{b_j C}{\tilde{\lambda_j}}\rvert^2}}\sum_{j=0}^{2^{n_b}-1}\frac{b_j C}{\tilde{\lambda_j}}\ket {u_j}\sum_{y=0}^{2^{n}-1} \ket{y}\ket{1}_a\nonumber\\
& = & \frac{C}{2^{n/2}\sqrt{\sum_{j=0}^{2^{n_b}-1}\lvert\frac{b_j C}{\lambda_j}\rvert^2}}\ket {x}\sum_{y=0}^{2^{n}-1} \ket{y}\ket{1}_a
\end{align}

The clock qubits and the b-register are now \emph{unentangled} and the b-register stores $\ket{x}$. By applying the Hadamard gate on the clock qubits, finally, we have

\begin{align}
\ket{\Psi_9}
& =  \frac{1}{\sqrt{\sum_{j=0}^{2^{n_b}-1}\lvert\frac{b_j C}{\lambda_j}\rvert^2}}\sum_{j=0}^{2^{n_b}-1}\frac{b_j C}{\lambda_j}\ket {u_j} \ket{0}^{\otimes n}\ket{1}_a\nonumber\\
& =  \frac{C}{\sqrt{\sum_{j=0}^{2^{n_b}-1}\lvert\frac{b_j C}{\lambda_j}\rvert^2}}\ket {x}_b \ket{0}_c^{\otimes n}\ket{1}_a
\end{align}

If $C$ is real and by using (\ref{eq7}),
\begin{align}
\ket{\Psi_9}
& =  \frac{1}{\sqrt{\sum_{j=0}^{2^{n_b}-1}\lvert\frac{b_j }{\lambda_j}\rvert^2}}\ket {x}_b \ket{0}_c^{\otimes n}\ket{1}_a\nonumber\\
& = \ket {x}_b \ket{0}_c^{\otimes n}\ket{1}_a
\end{align}

Here, the solution $\ket{x}$ (Eq.~\ref{xsolution}) is stored in the b-register successfully. 

\section{Numerical Example} \label{numerical_example}

We will present a numerical example and apply HHL to it step-by-step. The implementation is shown in Figure~\ref{HHL_Circuit}. Firstly, we will discuss how to implement the controlled-$U$ and ancilla qubit rotations. 

\subsection{Encoding Scheme}
In this example, the matrix $A$ and vector $\vec{b}$ are set to be

\begin{equation}
A
=
\begin{pmatrix}
1 & -\frac{1}{3} \\
-\frac{1}{3} & 1 \\
\end{pmatrix}
\end{equation}

\begin{equation}
\vec{b}
=
\begin{pmatrix}
0 \\
1 \\
\end{pmatrix}
\end{equation}

The eigenvectors of $A$ are $\vec{u_0}=\begin{pmatrix} \frac{-1}{\sqrt{2}} \\ \frac{-1}{\sqrt{2}} \end{pmatrix}$, $\vec{u_1}=\begin{pmatrix} \frac{-1}{\sqrt{2}} \\ \frac{1}{\sqrt{2}} \end{pmatrix}$ with eigenvalues $\lambda_0=\frac{2}{3}$ and $\lambda_1=\frac{4}{3}$ respectively. We need to using \emph{basis encoding} to encode the eigenvalues in the basis formed by the clock qubit and 2 qubits are needed by encoding $\lambda_0$ as $\ket{01}$ and $\lambda_1$ as $\ket{10}$ so that it maintains the ratio of $\lambda_1/\lambda_0=2$. This means $\tilde{\lambda_0}=1$ and $\tilde{\lambda_1}=2$ or in other words, $\tilde{\ket{\lambda_0}}=\ket{01}$ and $\tilde{\ket{\lambda_1}}=\ket{10}$. This gives a perfect encoding with $n=2$ (i.e. $N=4$). Therefore, $t$ is chosen to be $\frac{3\pi}{4}$ to achieve the encoding scheme since $\tilde{\lambda_j}=N\lambda_j t/2\pi$.

Since $\vec{b}$ is a 2-dimensional complex vector, it can be encoded using 1 qubit and, thus, $n_b=1$. 

The solution to the LSP is found to be

\begin{equation}\label{28}
\vec{x}
=
\begin{pmatrix}
\frac{3}{8} \\
\frac{9}{8} \\
\end{pmatrix}
\end{equation}
whereby, the ratio of $\lvert x_0 \rvert^2$ to $\lvert x_1\rvert^2$ is $1:9$.

\subsection{Controlled-U Implementation}
In reality, we expect the controlled-$U$ operation to be implemented by Hamiltonian simulation \cite{Dominic2007}. However, to understand the algorithm, we will derive the matrix for $U$ and then map this to the $Controlled-U$ gate used in IBM-Q directly. Since $n=2$, there are two operations needed, namely $U^{2^1}=U^2$ and $U^{2^0}=U$, controlled by $c_1$ and $c_0$, respectively.

In order to find the corresponding matrix for $U^2=e^{i2At}$ and $U=e^{iAt}$, we need to perform \emph{similarity transformation} on $i2At$ and $iAt$, exponentiate then, and transforms back to the original basis.

The transformation matrix from the original basis to the eigenvector basis is
\begin{align}
\label{eq:29}
V
& =  \begin{pmatrix} \vec{u_0} & \vec{u_1}\end{pmatrix}\nonumber\\
& = \begin{pmatrix} \frac{-1}{\sqrt{2}} & \frac{-1}{\sqrt{2}}\\ \frac{-1}{\sqrt{2}} & \frac{1}{\sqrt{2}}\end{pmatrix}
\end{align}

Since $V$ is real and symmetric,its conjugate, $V^\dagger$ equals itself.

The diagonalized $A$, i.e. expressed in the basis formed by $\vec{u_0}$ and $\vec{u_0}$, is 

\begin{align}
A_{diag}
& =  V^\dagger A V \nonumber\\
&=\begin{pmatrix}\frac{2}{3} & 0\\0 & \frac{4}{3}\end{pmatrix}
\end{align}

As it is diagonal, $U$ can be obtained by exponentiation of the elements accordingly.

\begin{align}
U_{diag}
& =  \begin{pmatrix}e^{i\lambda_0t} & 0\\0 & e^{i\lambda_1t}\end{pmatrix}\nonumber\\
& =  \begin{pmatrix}e^{i\pi/2} & 0\\0 & e^{i\pi}\end{pmatrix}\nonumber\\
& =  \begin{pmatrix}i & 0\\0 & -1\end{pmatrix}
\end{align}

where $t=3\pi/4$ as mentioned earlier is used. Also,

\begin{align}
U_{diag}^2
=U_{diag}U_{diag} =  \begin{pmatrix}-1 & 0\\0 & 1\end{pmatrix}
\end{align}

It is worth noting that both are naturally \emph{unitary} which is a requirement for a quantum operation.

To obtain $U$ and $U^2$ in the original basis, we need to apply similarity transformation again in the reverse direction,

\begin{align}
U
 &=  V U_{diag} V^{\dagger}\nonumber\\
& =  \begin{pmatrix} \frac{-1}{\sqrt{2}} & \frac{-1}{\sqrt{2}}\\ \frac{-1}{\sqrt{2}} & \frac{1}{\sqrt{2}}\end{pmatrix}\begin{pmatrix}i & 0\\0 & -1\end{pmatrix} \begin{pmatrix} \frac{-1}{\sqrt{2}} & \frac{-1}{\sqrt{2}}\\ \frac{-1}{\sqrt{2}} & \frac{1}{\sqrt{2}}\end{pmatrix}\nonumber\\
& = \frac{1}{2} \begin{pmatrix}-1+i & 1+i\\1+i & -1+i\end{pmatrix}
\end{align}

\begin{align}
U^2
& =  V U_{diag}^2 V^{\dagger}\nonumber\\
& =  \begin{pmatrix} \frac{-1}{\sqrt{2}} & \frac{-1}{\sqrt{2}}\\ \frac{-1}{\sqrt{2}} & \frac{1}{\sqrt{2}}\end{pmatrix}\begin{pmatrix}-1 & 0\\0 & 1\end{pmatrix} \begin{pmatrix} \frac{-1}{\sqrt{2}} & \frac{-1}{\sqrt{2}}\\ \frac{-1}{\sqrt{2}} & \frac{1}{\sqrt{2}}\end{pmatrix}\nonumber\\
& = \begin{pmatrix}0 & -1\\-1 & 0\end{pmatrix}
\end{align}

To implement $U$ and $U^2$, a 4-parameter arbitrary unitary gate with global phase \cite{Wong2022},  

\begin{align}
U
& =  \begin{pmatrix} e^{i\gamma}\cos(\theta/2)&-e^{i(\gamma+\lambda)}\sin(\theta/2)\\ e^{i(\gamma+\phi)}\sin(\theta/2)&e^{i(\gamma+\phi+\lambda)}\cos(\theta/2)
\end{pmatrix}
\end{align}

By choosing $\theta =\pi, \phi=\pi, \lambda=0, \gamma=0$, $U^2$ is implemented.

By choosing $\theta =\pi/2, \phi=-\pi/2, \lambda=\pi/2, \gamma=3\pi/4$, $U$ is implemented.

For the IQPE part, we also need to implement $U^{-1}$ and $U^{-2}$. Since in this example, ${(U^2)}^{-1}=U^2$, one can use the same set of parameters to implement $(U^2)^{-1}$.

However,

\begin{align}
U^{-1}
& = \frac{1}{2} \begin{pmatrix}-1-i & 1-i\\1-i & -1-i\end{pmatrix}
\end{align}

We need to choose $\theta =\pi/2, \phi=\pi/2, \lambda=-\pi/2, \gamma=-3\pi/4$ to implement $U^{-1}$.

The controlled version of matrix $U'$ can then be constructed using

\begin{align}
C-U'=I\otimes \ket{0}\bra{0}+U'\otimes \ket{1}\bra{1}
\label{controleq}
\end{align}

Note that in this equation, only the controlling clock bit and the b-register are included for simplicity. For example, 

\begin{align}
C-U^{-1}
&=\begin{pmatrix}1 & 0 \\0 & 1\end{pmatrix}\otimes
\begin{pmatrix} 1 & 0\\0 & 0\end{pmatrix}+
\frac{1}{2} \begin{pmatrix}-1-i & 1-i\\1-i & -1-i\end{pmatrix} \nonumber\\
&\otimes \begin{pmatrix} 0 & 0\\0 & 1\end{pmatrix}\nonumber\\
&=\begin{pmatrix}1 & 0 & 0 &0 \\0&0&0&0\\0&0&1&0\\0&0&0&0\end{pmatrix}+
\frac{1}{2}\begin{pmatrix}0 & 0 & 0 &0 \\0&-1-i&0&1-i\\0&0&0&0\\0&1-i&0&-1-i\end{pmatrix}\nonumber\\
&=\frac{1}{2}\begin{pmatrix}2 & 0 & 0 &0 \\0&-1-i&0&1-i\\0&0&2&0\\0&1-i&0&-1-i\end{pmatrix}
\end{align}

\subsection{Implementataion of the Controlled-Rotation of Ancilla Qubit}

The coefficients of $\ket{0}$ and $\ket{1}$ of the ancilla bit after rotation in Eq.~(\ref{ancilla_R}) are $\sqrt{1-\frac{C^2}{\tilde{\lambda_j}^2}}$ and $\frac{C}{\tilde{\lambda_j}}$, respectively. The sum of the square of the magnitudes of the coefficients is $1$ as required. This means also $C\leq\tilde{\lambda_j}$. Since the minimal $\tilde{\lambda_j}$ is $1$, we will set $C=1$ to maximize the probability of measuring $\ket{1}$ during the ancilla bit measurement.

The transformation of $\ket{0}_a$ to $\sqrt{1-\frac{1}{\tilde{\lambda_j}^2}}\ket{0}_a+\frac{1}{\tilde{\lambda_j}}\ket{1}_a$ is known to be equivalent to $RY(\theta)$ rotation, 

\begin{align}
RY(\theta) = \begin{pmatrix}\cos(\frac{\theta}{2}) & -\sin(\frac{\theta}{2})\\ \sin(\frac{\theta}{2}) & \cos(\frac{\theta}{2})\end{pmatrix}
\end{align}
with $\theta= 2\arcsin(\frac{1}{\tilde{\lambda_j}})$. One can check this by multiplying $RY(\theta)$ to $\ket{0}_a$. 

\begin{align}
RY(\theta)\ket{0}_a = \begin{pmatrix}\cos(\frac{\theta}{2}) & -\sin(\frac{\theta}{2})\\ \sin(\frac{\theta}{2}) & \cos(\frac{\theta}{2})\end{pmatrix}\begin{pmatrix}
1\\0
\end{pmatrix} \nonumber \\
=\cos(\frac{\theta}{2})\ket{0}_a+\sin(\frac{\theta}{2})\ket{1}_a
\end{align}

Therefore, we will establish a function to implement this rotation and this function only need to be valid when the input are the encoded eigenvalues because only encoded eigenvalues have zero magnitudes in the c-register as shown in (\ref{finalPsi4}). The function is defined as

\begin{align}
\label{arcsine}
\theta(c) =\theta(c_1c_0)= 2\arcsin(\frac{1}{c})
\end{align}

where $c$ is the value of the clock qubits and $c_1c_0$ is its binary form. 

Since only $\ket{\tilde{\lambda_j}}$ has non-zero amplitude in (\ref{finalPsi4}), we only need to set up (\ref{arcsine}) such that it is correct for $\ket{c}$=$\ket{01}$ and $\ket{10}$, namely

\begin{align}
\theta(1)=\theta(01)=2\arcsin(\frac{1}{1}) = \pi\\
\theta(2)=\theta(10)=2\arcsin(\frac{1}{2})=\frac{\pi}{3}
\end{align}

The following function can achieve the goal, 

\begin{align}
\theta(c)=\theta(c_1c_0)= \frac{\pi}{3}c_1+\pi c_0
\end{align}

Therefore, the controlled rotation can be implemented as

\begin{align}
\ket{1}\bra{1}\otimes I\otimes RY(\frac{\pi}{3})+\ket{0}\bra{0}\otimes I \otimes I + \nonumber\\
 I \otimes \ket{1}\bra{1} \otimes RY(\pi)+ I \otimes \ket{0}\bra{0}\otimes I 
\end{align}

where the operators operate on qubits $\ket{c_1}$, $\ket{c_0}$, and $\ket{a}$ from left to right, respectively.

\subsection{Quantum Circuit}

\begin{figure*}[h!]
\centering
\includegraphics[width=5in]{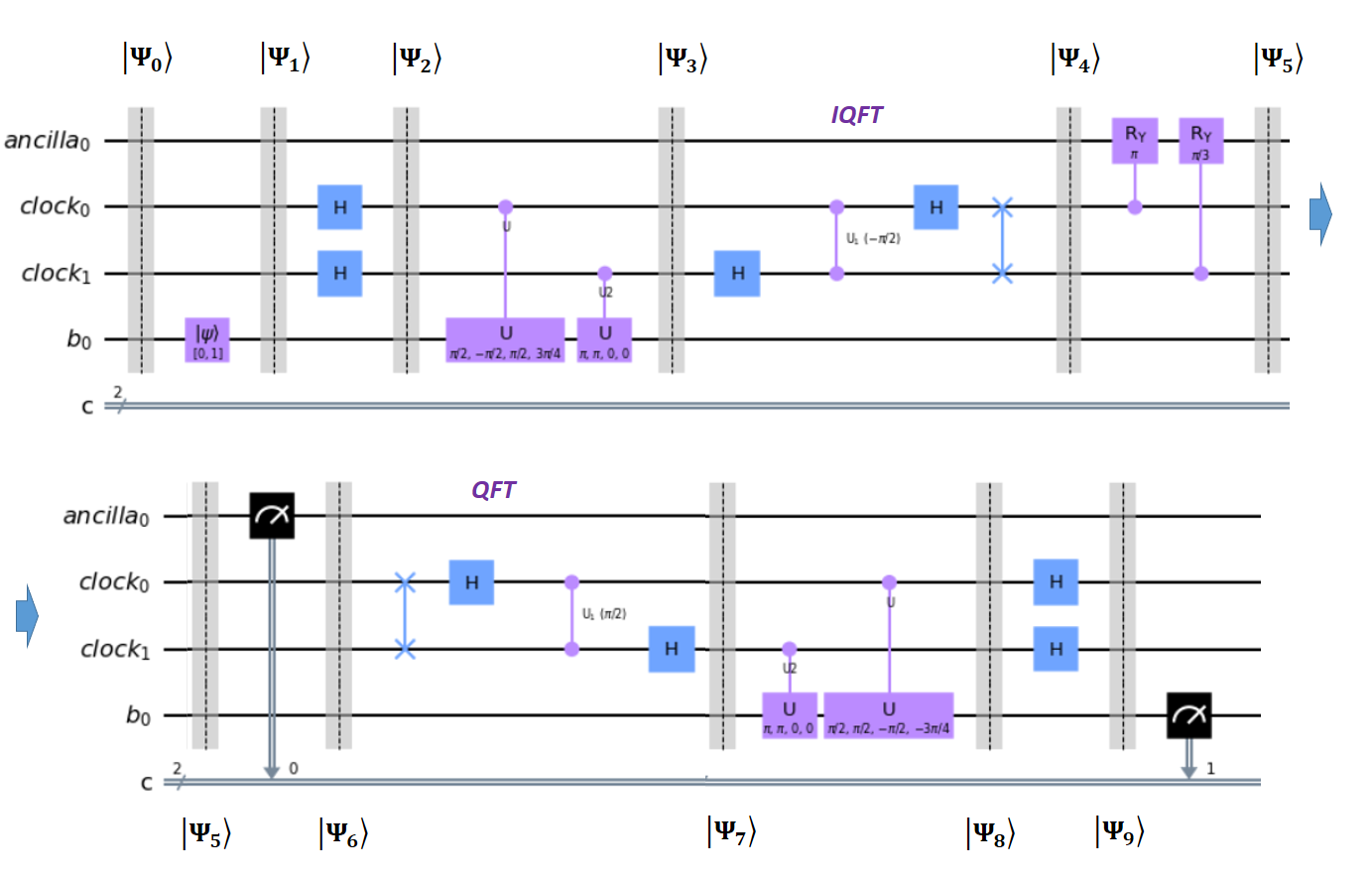}
\caption{The HHL circuit corresponding to the numerical example built to run in IBM-Q. This circuit is partitioned in the same way as the general HHL schematic in Figure~\ref{Schematic}.}
\label{HHL_Circuit}
\end{figure*}

An HHL circuit for the numerical example is then built and shown in Figure~\ref{HHL_Circuit}. We will then walk through the circuit using numerical substitution.

\subsection{Numerical Substitution}

The algorithm begins with
\begin{equation}
\ket {\Psi_0} = \ket{0}_b \otimes \ket{00}_c\otimes \ket{0}_a = \ket{0000}
\end{equation}
X-gate is then applied to convert $\ket{0}_b$ to $\ket{1}_b$ with
\begin{equation}
\label{eq:46}
\ket {\Psi_1} = X \otimes I\otimes I \ket{\Psi_0} = \ket{1000}
\end{equation}

After applying the Hadamard gates to create a superposition among the clock qubits,
\begin{align}
\label{eq:47}
\ket{\Psi_2}
& =  I\otimes H^{\otimes n}\otimes I \ket{\Psi_1}\nonumber\\
& =  \ket {1}\frac{1}{2^{\frac{2}{2}}}(\ket{0}+\ket{1})^{\otimes 2}\ket{0}\nonumber\\
& =  \frac{1}{2}(\ket{1000}+\ket{1010}+\ket{1100}+\ket{1110})
\end{align}

Before applying the CU3(controlled rotation of ancillary qubit) gates in the bra-ket notation, it will be convenient to perform a \emph{basis change} to the eigenvector basis of $A$. Since $\ket{1}=\frac{1}{\sqrt{2}}(-\ket{u_0}+\ket{u_1})$, we have $b_0=\frac{-1}{\sqrt{2}}$ and $b_1=\frac{1}{\sqrt{2}}$. Therefore,
\begin{align}
\ket{\Psi_2}
& =  \ket {1}\frac{1}{2}(\ket{000}+\ket{010}+\ket{100}+\ket{110})\nonumber\\
& =  \frac{1}{\sqrt{2}}(-\ket{u_0}+\ket{u_1})\frac{1}{2}(\ket{000}+\ket{010}+\ket{100}\nonumber\\
&+\ket{110})\nonumber\\
& =  \frac{1}{2\sqrt{2}}(-\ket{u_0}\ket{000}-\ket{u_0}\ket{010}-\ket{u_0}\ket{100}\nonumber
\\
&-\ket{u_0}\ket{110}+\ket{u_1}\ket{000}+\ket{u_1}\ket{010}\nonumber\\
&+\ket{u_1}\ket{100}+\ket{u_1}\ket{110})
\end{align}

In the controlled rotation operations, when the corresponding c-register is $\ket{k}_c$, a phase change of $\phi_j=k\lambda_j t/2\pi$ is added (i.e. multiplied by $e^{2\pi i \phi_j}$) for $\ket{u_j}$. Since $t=\frac{3\pi}{4}$, $\lambda_0=\frac{2}{3}$ and $\lambda_1=\frac{4}{3}$, we have
\begin{align}
\ket{\Psi_3}
& = \frac{1}{2\sqrt{2}}(-\ket{u_0}\ket{000}-e^{2\pi i \phi_0}\ket{u_0}\ket{010}\nonumber\\
&-e^{2\pi i 2\phi_0}\ket{u_0}\ket{100} 
-e^{2\pi i 3\phi_0}\ket{u_0}\ket{110}+\nonumber\\
&\ket{u_1}\ket{000}+e^{2\pi i \phi_1}\ket{u_1}\ket{010} \nonumber \\
&+e^{2\pi i 2\phi_1}\ket{u_1}\ket{100}+e^{2\pi i 3\phi_1}\ket{u_1}\ket{110})\nonumber\\
& =  \frac{1}{2\sqrt{2}}(-\ket{u_0}\ket{000}-e^{i\lambda_0 t}\ket{u_0}\ket{010}\nonumber\\
&-e^{i2\lambda_0 t}\ket{u_0}\ket{100} 
-e^{i3\lambda_0 t}\ket{u_0}\ket{110}+\nonumber\\
&\ket{u_1}\ket{000}+e^{i\lambda_1 t}\ket{u_1}\ket{010}\nonumber \\
&+e^{i2\lambda_1 t}\ket{u_1}\ket{100} +e^{i3\lambda_1 t}\ket{u_1}\ket{110})\nonumber\\
& =  \frac{1}{2\sqrt{2}}(-\ket{u_0}\ket{000}-e^{i\pi/2}\ket{u_0}\ket{010}-e^{i\pi}\ket{u_0}\ket{100} \nonumber\\
&-e^{i3\pi/2}\ket{u_0}\ket{110}+\ket{u_1}\ket{000}+e^{i\pi}\ket{u_1}\ket{010}\nonumber \\
&+e^{i2\pi}\ket{u_1}\ket{100} +e^{i3\pi}\ket{u_1}\ket{110})\nonumber\\
& =  \frac{1}{2\sqrt{2}}(-\ket{u_0}\ket{000}-i\ket{u_0}\ket{010}+\ket{u_0}\ket{100} \nonumber\\
&+i\ket{u_0}\ket{110}+\ket{u_1}\ket{000}-\ket{u_1}\ket{010}+\nonumber\\
&\ket{u_1}\ket{100} -\ket{u_1}\ket{110})
\end{align}

Before applying $IQFT$, the terms are regrouped for simplicity.
\begin{align}
\ket{\Psi_3}
& =  \frac{1}{2\sqrt{2}}((-\ket{u_0}+\ket{u_1})\ket{00}+(-i\ket{u_0}-\ket{u_1})\ket{01}\nonumber\\
& +(\ket{u_0} +\ket{u_1} )\ket{10} +(i\ket{u_0}-\ket{u_1})\ket{11})\ket{0}
\end{align}

Now apply $IQFT$ to the clock qubits, e.g.
\begin{align}
\textrm{IQFT}\ket{10}
& =  \textrm{IQFT}\ket{2}\nonumber\\
& =  \frac{1}{2^{2/2}} \sum_{y=0}^{2^{2}-1} e^{-2\pi i 2 y/4} \ket{y}\nonumber\\
& =  \frac{1}{2} (\ket{0}- \ket{1}+ \ket{2}- \ket{3}\nonumber\\
& =  \frac{1}{2} (\ket{00}- \ket{01}+ \ket{10}- \ket{11})
\label{e1}
\end{align}
Similarly,
\begin{align}
\textrm{IQFT}\ket{00}
\label{eq:53}
& =  \frac{1}{2} (\ket{00}+ \ket{01}+ \ket{10}+ \ket{11})
\end{align}
\begin{align}
\textrm{IQFT}\ket{01}
\label{eq:54}
& = \frac{1}{2} (\ket{00}- i\ket{01}- \ket{10}+i \ket{11})
\end{align}
\begin{align}
\textrm{IQFT}\ket{11}
\label{e4}
& =  \frac{1}{2} (\ket{00}+ i\ket{01}- \ket{10}-i \ket{11})
\end{align}
Therefore, applying $IQFT$ to $\ket{\Psi_3}$ and substituting Eq.~(\ref{e1}) to (\ref{e4}),
\begin{align}
\label{eq:56}
\ket{\Psi_4}
& =  \textrm{IQFT}\ket{\Psi_3}\nonumber\\
& =  \frac{1}{4\sqrt{2}}\nonumber\\
&((-\ket{u_0}+\ket{u_1})(\ket{00}+ \ket{01}+ \ket{10}+ \ket{11})+\nonumber\\
&(-i\ket{u_0}-\ket{u_1})(\ket{00}- i\ket{01}- \ket{10}+i \ket{11})+\nonumber\\
&(\ket{u_0} +\ket{u_1} ) (\ket{00}- \ket{01}+ \ket{10}- \ket{11}) +\nonumber\\
&(i\ket{u_0}-\ket{u_1}) (\ket{00}+ i\ket{01}- \ket{10}-i \ket{11}))\ket{0}\nonumber\\
&= \frac{1}{\sqrt{2}}(-\ket{u_0}\ket{01}+\ket{u_1} \ket{10})\ket{0}
\end{align}

It can be seen that after $IQFT$, the eigenvalues are encoded in the clock qubits as $\ket{01}$ and $\ket{11}$ with non-zero amplitudes due \emph{constructive interference}. $b_0=\frac{-1}{\sqrt{2}}$ and $b_1=\frac{1}{\sqrt{2}}$. We clearly see the entanglement between the b-register and the c-register that $\ket{u_0}$ goes with $\ket{01}$ and $\ket{u_1}$ goes with $\ket{11}$.

After performing the ancilla qubit rotation, 
\begin{align}
\ket{\Psi_5}
&=\sum_{j=0}^{2^{1}-1}b_j\ket {u_j}\ket{\tilde{\lambda_j}}(\sqrt{1-\frac{C^2}{\tilde{\lambda_j}^2}}\ket{0}+\frac{C}{\tilde{\lambda_j}}\ket{1})\nonumber\\
&=-\frac{1}{\sqrt{2}}\ket{u_0}\ket{01}(\sqrt{1-\frac{1}{1^2}}\ket{0}+\frac{1}{1}\ket{1})+\nonumber\\
&\frac{1}{\sqrt{2}}\ket{u_1} \ket{10}(\sqrt{1-\frac{1}{2^2}}\ket{0}+\frac{1}{2}\ket{1})
\end{align}

If the measurement of the ancilla bit is $\ket{1}$,
\begin{align}
\ket{\Psi_6}
&=\sqrt{\frac{8}{5}}(-\frac{1}{\sqrt{2}}\ket{u_0}\ket{01}\ket{1}\nonumber \\
&+\frac{1}{2\sqrt{2}}\ket{u_1} \ket{10}\ket{1})
\label{57}
\end{align}

Applying \textrm{QFT} to the encoded eigenvalues, we have
\begin{align}
\label{e5}
\textrm{QFT}\ket{10}
& =  \textrm{QFT}\ket{2}\nonumber\\
& =  \frac{1}{2^{2/2}} \sum_{y=0}^{2^{2}-1} e^{2\pi i 2 y/4} \ket{y}\nonumber\\
& =  \frac{1}{2} (\ket{00}- \ket{01}+ \ket{10}- \ket{11})
\end{align}
\begin{align}
\label{e6}
\textrm{QFT}\ket{01}
& =  \textrm{QFT}\ket{1}\nonumber\\
& =  \frac{1}{2} (\ket{00}+i \ket{01}- \ket{10}-i \ket{11})
\end{align}

Therefore, applying QFT to $\ket{\Psi_6}$ and substituting Eq.~(\ref{e5}) to (\ref{e6}), we obtain
\begin{align}
\label{eq:61}
\ket{\Psi_7}
&=\sqrt{\frac{8}{5}}(-\frac{1}{\sqrt{2}}\ket{u_0}\frac{1}{2} (\ket{00}+i \ket{01}- \ket{10}\nonumber\\
&-i \ket{11})\ket{1}+\nonumber\frac{1}{2\sqrt{2}}\ket{u_1} \frac{1}{2} (\ket{00}- \ket{01}\nonumber\\
&+ \ket{10}- \ket{11}))\ket{1}
\end{align}

For the controlled rotation, the state is multiplied by $e^{-i\lambda_j t}$ and $e^{-i2\lambda_j t}$ if $c_0=1$ and $c_1=1$, respectively. Since $e^{-i\lambda_0 t}=-i$, $e^{-i2\lambda_0 t}=-1$, $e^{-i\lambda_1 t}=-1$, $e^{-i2\lambda_1 t}=1$, and $Nt/2\pi$=$3/2$
\begin{align}
\ket{\Psi_8}
&=\sqrt{\frac{8}{5}}(-\frac{1}{\sqrt{2}}\ket{u_0}\frac{1}{2} (\ket{00}+\ket{01}+ \ket{10}+ \ket{11})\ket{1}\nonumber\\
&+\frac{1}{2\sqrt{2}}\ket{u_1} \frac{1}{2} (\ket{00}+\ket{01}+ \ket{10}+\ket{11}))\ket{1}\nonumber\\
&=\frac{1}{2}\sqrt{\frac{8}{5}}(-\frac{1}{\sqrt{2}}\ket{u_0}+\frac{1}{2\sqrt{2}}\ket{u_1}) (\ket{00}+\ket{01} \nonumber\\
& +\ket{10}+\ket{11})\ket{1}\nonumber\\
&=\frac{1}{2}(\frac{2}{3})\sqrt{\frac{8}{5}}(-\frac{1}{\frac{2}{3}\sqrt{2}}\ket{u_0} +\frac{1}{\frac{4}{3}\sqrt{2}}\ket{u_1}) (\ket{00}+\ket{01}\nonumber\\
&+ \ket{10}+\ket{11})\ket{1}
\end{align}

Finally, by applying Hadamard gate to the clock qubits,

\begin{align}
\label{final}
\ket{\Psi_9}
&=\frac{2}{3}\sqrt{\frac{8}{5}}(-\frac{1}{\frac{2}{3}\sqrt{2}}\ket{u_0}\nonumber\\
&+\frac{1}{\frac{4}{3}\sqrt{2}}\ket{u_1}) \ket{00}\ket{1}
\end{align}

It can be verified that $\ket{\Psi_9}$ is a normalized vector as it should be because every operation in the HHL circuit is unitary and preserves the norm.

Equation ({\ref{final}}) can be simplified by substituting $\ket{u_0}=\frac{-1}{\sqrt{2}}\ket{0}+ \frac{-1}{\sqrt{2}}\ket{1}$ and $\ket{u_1}=\frac{-1}{\sqrt{2}}\ket{0}+ \frac{1}{\sqrt{2}}\ket{1}$. We obtain,

\begin{align}
\ket{\Psi_9}
&=\frac{1}{2}\sqrt{\frac{2}{5}}(\ket{0} +3\ket{1}) \ket{00}\ket{1}
\label{eq67}
\end{align}

The probability ratio of obtaining $\ket{0}$ and $\ket{1}$ when b-register is measured is thus $1:9$ as expected.

\subsection{Simulation Results}

Matlab code implementing the numerical example using matrix approach is created and available at \cite{matlab}. In the Matlab code, measurement is not performed (i.e. not partical tracing of the matrix). $\Psi_9$ is found to be,
\begin{equation}
\ket{\psi_9}
=
\begin{pmatrix}
-0.4330\\
0.2500\\
0.0000\\
-0.0000\\
0.0000\\
-0.0000\\
0.0000\\
0.0000\\
0.4330\\
0.7500\\
-0.0000\\
0.0000\\
-0.0000\\
0.0000\\
-0.0000\\
0.0000\\
\end{pmatrix}
\end{equation}
Since $\ket{0}_c$ are discarded during the measurement step, only $\ket{0001}$ and $\ket{1001}$ are left. Their amplitude ratio is $0.25^2:0.75^2=1:9$ as expected.

\begin{figure}[h!]
\centering
\includegraphics[width=3.5in]{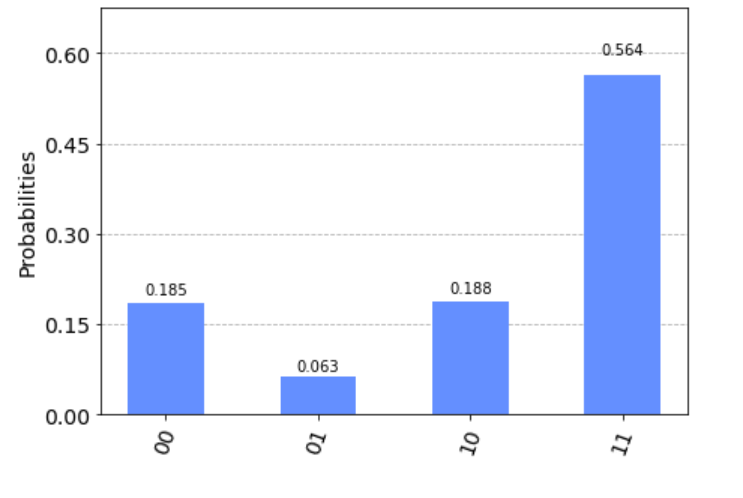}
\caption{Simulation result of the circuit in Figure~\ref{HHL_Circuit} using $IBM-Q$. Only the MSB $\ket{\;}_b$, and the LSB $\ket{\;}_a$ are measured.}
\label{simulation_result}
\end{figure}

The circuit in Fig.~\ref{HHL_Circuit} is also simulated in the IBM-Q system (code available at \cite{matlab}). Since only the b-register and the ancilla qubit are measured, there are only four possible outputs as shown in Figure~\ref{simulation_result}. Again, only $\ket{1}_a$ should be considered. The ratio of the measurement probability of $\ket{0}_b\ket{1}_a$ to $\ket{1}_b\ket{1}_a$ is $0.063:0.564=1:8.95$, which is close to the expected value.

\begin{figure}[h!]
\centering
\includegraphics[width=3.5in]{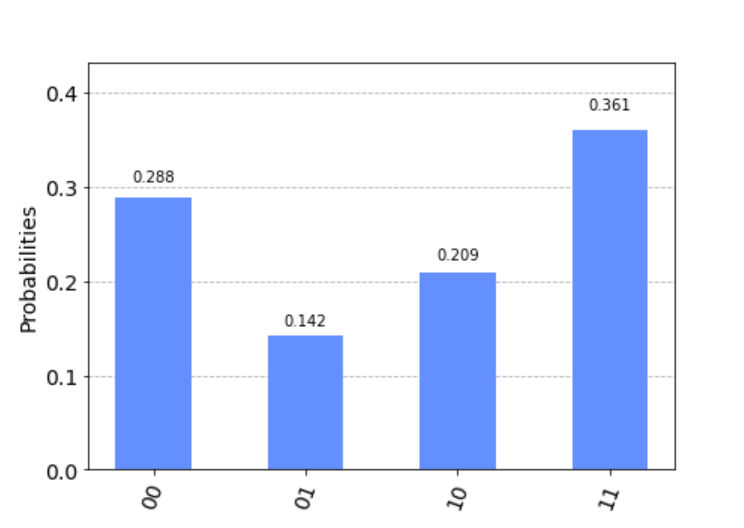}
\caption{Hardware result of the circuit in Figure~\ref{HHL_Circuit} run in machine ibmq\_santiago. Only the MSB, $\ket{\;}_b$ and the LSB, $\ket{\;}_a$ are measured.}
\label{hardware_result}
\end{figure}

On the other hand, due to the imperfection and noise in a real quantum computer, the hardware execution of the same circuit does not give a satisfactory result (Figure~\ref{hardware_result}). The ratio of the measurement probability of $\ket{0}_b\ket{1}_a$ to $\ket{1}_b\ket{1}_a$ is only $0.142^2:0.361^2=1:2.54$.

\section{Conclusion}
In this paper, we presented the HHL algorithm through a step-by-step walkthrough of the derivation. A numerical example is also presented in the bra-ket notation. The numerical example echos the analytical derivation to help students understand how qubits evolve in this important and relatively complex algorithm. A Matlab code corresponding to the numerical example is constructed to help understand the algorithm from the matrix point of view. Qiskit circuit of the corresponding circuit which can be simulated in IBM-Q and run on their quantum computing hardware is also available. Through this self-contained and step-by-step walkthrough, the basic concepts in quantum computing are reinforced.

\section{APPENDIX}
\label{appendix}
\numberwithin{equation}{section}

\subsubsection{Hermitian matrix } \label{A_Hermitian}
A Hermitian matrix is a matrix that equals to its adjoint matrix (transpose followed by complex conjugation). That is, if $A$ is a Hermitian matrix, then it is defined as,
\begin{align}
A =A^\dagger = (A^T)^*
\end{align}
where $A^T$ is the transpose of $A$. 

In this paper, the matrix, $A$, in the LPS to be solved is assumed to be Hermitian.

Another example is in (\ref{eq:29}), where $V$ is Hermitian.
\begin{align*}
V = \begin{pmatrix} \vec{u_0} & \vec{u_1}\end{pmatrix}\nonumber\\
= \begin{pmatrix} \frac{-1}{\sqrt{2}} & \frac{-1}{\sqrt{2}}\\ \frac{-1}{\sqrt{2}} & \frac{1}{\sqrt{2}}\end{pmatrix}
\tag{\ref{eq:29}}
\end{align*}

\subsubsection{Bra-ket Notation} \label{Bra-Ket}
Bra-ket notation is commonly used in quantum mechanics. A vector $\vec{v}$ is represented as $\ket{v}$ in its ket form. The bra form of the vectors forms a dual space to the space of the kets. The bra form of $\vec{v}$ is $\bra{v}$.

In matrix representation, ket is the complex conjugate transpose of bra and vice versa. For example, if $\ket{v}=\begin{pmatrix}
1\\-i\end{pmatrix}$, then $\bra{v}=\begin{pmatrix}
1&i.
\end{pmatrix}$

\subsubsection{Superposition} \label{Superposition} Superposition or Quantum Superposition is a quantum state which is the linear combination of two or more basis states. For example, a superposition state can be $\ket{v} = c_1 \ket{1} + c_2 \ket{0}$ , where $c_1$ and $c_2$ are complex number and $\ket{1}$ and $\ket{0}$ are basis states. A Hadamard gate is a gate commonly used to create a superposition state (Appendix~\ref{hadamard}). 

\subsubsection{Basis Transformation and Quantum Gate} \label{Basis_Transform}
In quantum computing, we only care about the basis transformation due to rotation in the hyperspace. The transformation is equivalent to the multiplication of the basis vectors by a unitary matrix, $U$, which is the transformation matrix. All quantum gates can be defined as how the basis vectors are transformed from the initial basis vector to the final basis vectors. Usually, a quantum gate rotates a basis state into another basis state (e.g. the NOT gate) and has its classical counterpart. But there are some gates that rotate a basis state to a superposition of two or more basis states. Such gates have no classical counterparts. For example, a Hadamard gate defines how an initial basis vector is rotated to an equal superposition of two basis vectors (Appendix~\ref{hadamard}).

\subsubsection{Hadamard Gate} \label{hadamard}

The Hadamard gate is a quantum gate that does not have a classical counterpart. It rotates the basis state to create an equal superposition of the basis states. For a 1-qubit case, this means it has equal probability (i.e. $\frac{1}{2}$) of measuring $\ket{0}$ and $\ket{1}$.

The matrix form of the Hadamard gate is,
\begin{align}
\frac{1}{\sqrt{2}}
\begin{pmatrix}
  1 & 1 \\
  1 & -1 \\
\end{pmatrix}   
\end{align}
When it is applied on the basis state $\ket{1}$, which is 
$
\begin{pmatrix}
  0\\
  1\\
\end{pmatrix}
$
in matrix form, we have,
\begin{align}
=
\frac{1}{\sqrt{2}}
\begin{pmatrix}
  1 & 1 \\
  1 & -1 \\
\end{pmatrix}
\begin{pmatrix}
  0\\
  1\\
\end{pmatrix}
\\
=
\frac{1}{\sqrt{2}}
\begin{pmatrix}
  1\\
 -1\\
\end{pmatrix}
\end{align}

which can also be represented in bra-ket form as,
\begin{align}
\frac{\ket{0}-\ket{1}}{\sqrt{2}}
\end{align}

 In this paper, Hadamard gates are applied in clock qubit to create superposition from $\ket{\psi_1}$ to $\ket{\psi_2}$. For example, in (\ref{eq:9-10}),
\begin{align*}
\ket{\Psi_2}
= I^{\otimes n_b}\otimes H^{\otimes n} \otimes I\ket{\Psi_1}
\tag{\ref{eq:9-10}} 
\end{align*}
where $\ket{\psi_2}$ is obtained by applying tensor product of identity gates and an $n$-qubit Hadamard gate to $\ket{\psi_1}$. The identity gates are applied to the b-register and the ancilla qubit while the Hadamard gate is applied to the clock qubits. In this equation, the $n$-qubit Hadamard gate is represented as $H^{\otimes n}$, i.e. the tensor product of $n$ 1-qubit Hadamard gates.

\subsubsection{Entanglement} \label{entanglement}
Entanglement refers to the quantum state of a 2- or more-qubit system that cannot be expressed as a tensor product of the individual qubit. This is an important feature that quantum computing uses often. As an example,

\begin{align}
 \ket{\Psi}
= \frac{1}{\sqrt{2}}(\ket{00}+\ket{11})
\end{align}
is an entangled state. It cannot be expressed as a tensor product of two individual qubit states. 

In this paper, after the ancilla bit rotation, we have
\begin{align*}
\ket{\Psi_6}
=\sqrt{\frac{8}{5}}(-\frac{1}{\sqrt{2}}\ket{u_0}\ket{01}\ket{1}+\frac{1}{2\sqrt{2}}\ket{u_1} \ket{10}\ket{1})
\tag{\ref{57}}
\end{align*}
where the b-register and the c-register are entangled and $\ket{u_0}$ ($\ket{u_1}$) always appears with $\ket{01}$ ($\ket{10}$) after the measurement. 

If the b-register \emph{were} not entangled with the c-register, we have 

\begin{align}
\ket{\Psi_6}
=\sqrt{\frac{8}{5}}(-\frac{1}{\sqrt{2}}\ket{u_0}+\frac{1}{2\sqrt{2}}\ket{u_1})
\end{align}

By substituting $\ket{u_0}=\frac{-1}{\sqrt{2}}\ket{0}+ \frac{-1}{\sqrt{2}}\ket{1}$ and $\ket{u_1}=\frac{-1}{\sqrt{2}}\ket{0}+ \frac{1}{\sqrt{2}}\ket{1}$ and after simplification, we have

\begin{align}
\ket{\Psi_6}
&=\sqrt{\frac{8}{5}}(-\frac{1}{\sqrt{2}}(\frac{-1}{\sqrt{2}}\ket{0}+ \frac{-1}{\sqrt{2}}\ket{1})+\frac{1}{2\sqrt{2}}(\frac{-1}{\sqrt{2}}\ket{0}+\nonumber\\
&\frac{1}{\sqrt{2}}\ket{1}))\nonumber\\
&=\sqrt{\frac{8}{5}}(\frac{1}{4}\ket{0}+\frac{3}{4}\ket{1})
\end{align}

This is the same as (\ref{eq67}). The probability of measuring $\ket{0}$ and $\ket{1}$ has the ratio of 1:9 as expected. 

However, when there is entanglement, the probability of measuring $\ket{0}$ and $\ket{1}$ would not be 1:9 because the previous grouping is impossible.

\begin{align}
\ket{\Psi_6}
&=\sqrt{\frac{8}{5}}(-\frac{1}{\sqrt{2}}(\frac{-1}{\sqrt{2}}\ket{0}+ \frac{-1}{\sqrt{2}}\ket{1})\ket{01}+\frac{1}{2\sqrt{2}}(\frac{-1}{\sqrt{2}}\ket{0} \nonumber \\
&+ \frac{1}{\sqrt{2}}\ket{1}))\ket{10}
\end{align}

\subsubsection{Controlled Operation} \label{Controlled}
Controlled operation requires more than one qubit. For a 2-qubit controlled gate, an operation is applied to a qubit (the target qubit), if the value of the controlling qubit is 1 in the basis vector.

For example, in Figure~\ref{QPE-controlled}, $\ket{b}$ is the target qubit and $\ket{c_{n-1}}$ is the controlling qubit. The operation of $U^{2^{n-1}}$ is applied to $\ket{b}$ only if $\ket{c_{n-1}}$ is 1 in the basis state (e.g. $\ket{bc_{n-1}\cdots}=\ket{01\cdots}$. 

In general, the controlled version of a unitary gate, $U'$, can be implemented using the following equation if the LSB is the controlling qubit.

\begin{equation}
C-U'=I\otimes \ket{0}\bra{0}+U'\otimes \ket{1}\bra{1}
\tag{\ref{controleq}}
\end{equation}
which literally means that if the controlling qubit is 0, Identity gate is applied to the target qubit (MSB). Otherise, $U'$ is applied to the target qubit.

\subsubsection{Eigenvalue and Eigenvector} \label{A_Eigenvalues}

When a non-zero $n\times n$ matrix $A$ is applied to an $n$-dimensional vector $\vec{V}$ and has the following relationship,
\begin{align}
A  \vec{V} = \lambda   \vec{V}
\label{eigenvector_def}
\end{align}   
where $\lambda$ is a scalar, then, by definition, $\vec{V}$ and $\lambda$ are the eigenvector and eigenvalue of $A$, respectively. This is similar to (\ref{eq:3}), where the matrix $A$ is expressed as a linear combination of the outer products of its eigenvectors, $\ket{u_i} \bra{u_i}$.
\begin{align*}
 A=\sum_{i=0}^{2^{n_b}-1} \lambda_i \ket{u_i} \bra{u_i}
 \tag{\ref{eq:3}}
\end{align*}

This can be checked by applying $A$ to its eigenvector $\ket{u_j}$,
\begin{align}
 A\ket{u_j}&=\sum_{i=0}^{2^{n_b}-1} \lambda_i \ket{u_i} \bra{u_i} \ket{u_j}\\\nonumber
 &=\sum_{i=0}^{2^{n_b}-1} \lambda_i \ket{u_i} \delta_{ij}\\\nonumber
 &=\lambda_j \ket{u_j} 
\end{align}
which meets the definition in Eq.~({\ref{eigenvector_def}}).

\subsubsection{Different Types of Encoding} \label{A_Encoding}
The three common types of encodings are explained here.
\begin{itemize}
     \item Basis Encoding- Basis encoding converts classical information such as numbers or matrix to quantum information in the form of basis states. For example,
     \begin{align}
        && x = 2 &&
         \underrightarrow{\textrm{binary}}
         &&10&&
         \underrightarrow{\textrm{quantum state}}
         &&
         \ket{10}
         &&\nonumber
     \end{align}
     \begin{gather}
         x = 
         \begin{pmatrix}
         2\\
         3
         \end{pmatrix}
         \underrightarrow{\textrm{binary}}
         \begin{pmatrix}
          10\\
          11
         \end{pmatrix}
         \underrightarrow{\textrm{quantum state}}
         \ket{1011}
     \end{gather}
    \item Amplitude Encoding- Amplitude encoding encodes the information as the coefficients of the basis vectors. For example, for 
    \begin{align}
    \vec{v}
    =
    \begin{pmatrix}
    v_0\\
    v_1
    \end{pmatrix}
    \end{align}
    which is assumed to be normalized ($|v| = 1$), it can be encoded as in the following quantum state,
    \begin{align}
        \ket{v}
        =
        v_0
        \ket{0}
        + v_1
        \ket{1}
    \end{align}
    
    where $v_0$ and $v_1$ become the coefficients of the basis states, $\ket{0}$ and $\ket{0}$, respectively. In the main text, Eq.~(\ref{b-representation}) shows how the values of the components of vector $\ket{b}$ are encoded using amplitude encoding.
    
    \item Hamiltonian Encoding- One type of the Hamiltonian encoding is to encode the matrix as the Hamiltonian in a unitary gate. For example, in this paper, Eq.~(\ref{eq:15}) shows that
        \begin{align*}
            U=e^{iAt}
            \tag{\ref{eq:15}}
        \end{align*}
        where it encodes matrix $A$ as the Hamiltonian of the unitary gate $U$. Matrix $A$ needs to be Hermitian as it is used to represent the Hamiltonian (the energy) of the system. However, $A$ does not need to be unitary and $U$ will be unitary due to its definition in (\ref{eq:15}).
 \end{itemize}
\subsubsection{Discrete Fourier Transform (DFT)} \label{A_DFT} The discrete Fourier Transform (DFT) transforms an $N$-dimensional vector $\vec{X}$ to another $N$-dimensional vector $\vec{Y}$. The transformation matrix $\Omega$ contains the powers of the $N$-th root of unity, $\omega=e^{i2\pi/N}$. The transformation is represented as,
\begin{align}
\vec{Y}&=&\Omega\vec{X}\nonumber\\
\begin{pmatrix}y_0\\y_1\\\vdots\\y_{N-1}\end{pmatrix}
&=\frac{1}{\sqrt{N}}
\begin{pmatrix}\omega^{-0\cdot 0}& \cdots&\omega^{-0\cdot (N-1)}\\
\omega^{-1\cdot 0}& \cdots&\omega^{-1\cdot (N-1)}\\
\vdots&\ddots&\vdots\\
\omega^{-(N-1)\cdot 0}&\cdots&\omega^{-(N-1)\cdot (N-1)}\end{pmatrix}\nonumber\\
&\cdot \begin{pmatrix}x_0\\x_1\\\vdots\\x_{N-1}\end{pmatrix}
\label{eq_DFT}
\end{align}

\subsubsection{Inverse Quantum Fourier Transform (IQFT) and Quantum Fourier Transform(QFT)} \label{A_QFT}
Mathematically, IQFT is similar to DFT (See Appendix~\ref{A_DFT}). The transformation matrix, $U_I$, is $N\times N$ for an $N$-dimensional Hilbert space. Therefore, $N=2^n$ for an $n$-qubit system. Eq.~(\ref{eq_DFT}) becomes
\begin{align}
\ket{Y}
= U_I
\ket{X}
\end{align}

\noindent and $U_I$ has the same expression as $\Omega$ in DFT.
\begin{align}
U_{I}=\frac{1}{\sqrt{N}}
\begin{pmatrix}\omega^{-0\cdot 0}& \cdots&\omega^{-0\cdot (N-1)}\\
\omega^{-1\cdot 0}& \cdots&\omega^{-1\cdot (N-1)}\\
\vdots&\ddots&\vdots\\
\omega^{-(N-1)\cdot 0}&\cdots&\omega^{-(N-1)\cdot (N-1)}\end{pmatrix}
\label{IQFTMatrix}
\end{align}

 \emph{Note that in some literature, e.g. \cite{Wong2022}, this form of IQFT is called QFT}. $\ket{X}$ and $\ket{Y}$ are the quantum states/vectors in the $N$-dimensional Hilbert space. IQFT can be treated as the rotation of $\ket{X}$ to $\ket{Y}$.

If $\ket{X}$ is a basis vector $\ket{k}$, applying IQFT to $\ket{k}$ using Eq.~(\ref{IQFTMatrix}), we have

\begin{align}
U_I\ket{k}=\frac{1}{\sqrt{N}}\sum_{j=0}^{N-1}\omega^{-jk}\ket{j}
\label{25_13}
\end{align}

This is the equation used often in this paper. It tells us that by applying IQFT to a basis vector, the basis vector is rotated to a superposition of all basis vectors weighted by the powers of the $N$-th root of unity.

For example, in (\ref{e1}) in the main text,
\begin{align*}
\textrm{IQFT}\ket{10}
& =  \textrm{IQFT}\ket{2}\nonumber\\
& =  \frac{1}{2^{2/2}} \sum_{y=0}^{2^{2}-1} e^{-2\pi i 2 y/4} \ket{y}\nonumber\\
& =  \frac{1}{2} (\ket{0}- \ket{1}+ \ket{2}- \ket{3}\nonumber\\
& =  \frac{1}{2} (\ket{00}- \ket{01}+ \ket{10}- \ket{11}) \tag{\ref{e1}}
\end{align*}
where $N=2^2=4$, $k=2$, $j=y$ in (\ref{25_13}) is used. The basis state $\ket{10}$ becomes a superposition of all other basis states after $IQFT$.

Another more complex example is the general equation,in (\ref{eq:13}), for the $IQFT$ in Figure~\ref{Schematic}.

\begin{align*}
\ket{\Psi_4}
&=  \ket {b} \textrm{IQFT}(\frac{1}{2^{\frac{n}{2}}} \sum_{k=0}^{2^{n}-1} e^{2\pi i \phi k}\ket{k})\ket{0}_a\nonumber\\
&=  \ket {b} \frac{1}{2^{\frac{n}{2}}} \sum_{k=0}^{2^{n}-1} e^{2\pi i \phi k}(\textrm{IQFT}\ket{k})\ket{0}_a\nonumber\\
&=  \ket {b} \frac{1}{2^{n}}\sum_{k=0}^{2^{n}-1}e^{2\pi i \phi k}(\sum_{y=0}^{2^{n}-1} e^{-2\pi i y k/N} \ket{y})\ket{0}_a\nonumber\\
&= \frac{1}{2^{n}}\ket {b}\sum_{y=0}^{2^{n}-1} \sum_{k=0}^{2^{n}-1} e^{2\pi i k(\phi -y/N)}\ket{y}\ket{0}_a
\tag{\ref{eq:13}}
\end{align*}

Here, $IQFT$ is applied to the c-register which is a superposition of basis states, $\ket{k}$. Using the distribution law of matrix operations, $IQFT$ is applied to individual $\ket{k}$ and (\ref{25_13}) is used with $y=j$ and $N=2^n$.

QFT is the inverse of $IQFT$ and can be treated as the rotation of the basis. The rotation matrix is given by

\begin{align}
U_Q=
\frac{1}{\sqrt{N}}
\begin{pmatrix}1&1&\cdots&1\\
1& \omega^{1}& \cdots&\omega^{(N-1)}\\
1& \omega^{2}& \cdots&\omega^{2(N-1)}\\
\vdots&\vdots&\ddots&\vdots\\
1&\omega^{(N-1)}&\cdots&\omega^{(N-1)(N-1)}\end{pmatrix}
\label{eq_QFT}
\end{align}

Equivalently,
\begin{align}
U_Q\ket{k}=\frac{1}{\sqrt{N}}\sum_{j=0}^{N-1}\omega^{jk}\ket{j}
\label{QFT_eq}
\end{align}

It can be shown that $U_I=U_Q^{-1}$ or $U_IU_Q=I$.

\subsubsection{Implementation of $QFT$ and $IQFT$} \label{A_implement_QFT}
\begin{figure}[h!]
\centering
\includegraphics[height=0.8in,width=2in]{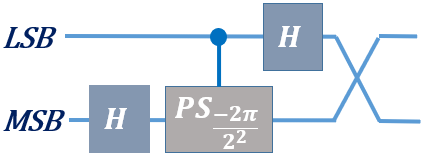}
\caption{ Implementation of a 2\-qubit inverse quantum Fourier transformation.}
\label{2_qbit_IQFT.PNG}
\end{figure}

$QFT$ and $IQFT$ are constructed using Hadamard gates, controlled phase shift gates, and $SWAP$ gates. Readers may refer to other sources for the details (e.g. \cite{Wong2022}). Here, we show the circuit of a 2-qubit $IQFT$ gate (Fig.~\ref{2_qbit_IQFT.PNG}).

In general, the phase shift angle is $\phi = \frac{-2\pi}{2^r}$, $r-1$ is the distance between the controlling qubit and the target qubit. For the 2-qubit $IQFT$ case, there is only one controlled phase shift gate and $r=2$ and this results in the phase $\phi = \frac{-\pi}{2}$.

For QFT, the circuit is the same as the $IQFT$, but the phase shift is negated, i.e. $\phi = \frac{2\pi}{2^r}$. This can be appreciated by the fact that the elements in the $IQFT$ and $QFT$ have opposition signs in (\ref{IQFTMatrix}) and (\ref{eq_QFT}), respectively.

\subsubsection{Gaussian elimination method} \label{A_Gaussian}
Here, Gaussian elimination is demonstrated by solving (\ref{tosolve}) using the numerical example in Section~\ref{numerical_example}. 
\begin{align*}
    A \vec{x} = \vec{b}
    \tag{\ref{tosolve}}
\end{align*}
\begin{align}
     \label{eq:95}
     \begin{pmatrix}
    1 & \frac{-1}{3}\\
    \frac{-1}{3} & 1
    \end{pmatrix}
        \begin{pmatrix}
    x_0\\
    x_1\
    \end{pmatrix}   
 =
    \begin{pmatrix}
    0\\
    1
    \end{pmatrix}   
    \end{align}
which is rewritten as an augmented matrix followed by Gaussian method of Elimination to solve for $x_0$ and $x_1$,
\begin{align*}
 \begin{bmatrix}
  1 & \frac{-1}{3} & 0 \\
    \frac{-1}{3} & 1 & 1
 \end{bmatrix}   
 \underrightarrow{\textrm{Row2} = 3\times \textrm{Row2} +\textrm{Row1}}   
 \begin{bmatrix}
  1 & \frac{-1}{3} & 0 \\
  0 & \frac{8}{3} & 3
 \end{bmatrix}
\end{align*}
\begin{align*}
 \underrightarrow{\textrm{Row2}= \frac{3}{8}\times \textrm{Row2}}
 \begin{bmatrix}
  3 & -1 & 0 \\
  0 & 1 & \frac{9}{8}
  \end{bmatrix} 
\end{align*}
 \begin{align*}
     \begin{bmatrix}
      3 & -1 & 0 \\
      0 & 1 & \frac{9}{8}
     \end{bmatrix}
     \underrightarrow{\textrm{Row1}=\frac{1}{3}\times \textrm{Row1} +\textrm{Row2}}
     \begin{bmatrix}
      1 & 0 & \frac{3}{8}\\
      0 & 1 & \frac{9}{8}
     \end{bmatrix}
 \end{align*}
 This the solution $\vec{x}$ is
 \begin{align*}
 \begin{pmatrix}
 x_0\\
 x_1
 \end{pmatrix}
 =
 \begin{pmatrix}
 \frac{3}{8}\\
 \frac{9}{8}
 \end{pmatrix}
 \end{align*}
 
 The complexity of Gaussian Elimination is $O(N^3)$. This is much slower than the classical conjugate gradient method (Appendix~\ref{A_conjugate}), to which HHL is compared.
 
\subsubsection{Conjugate Gradient Method} \label{A_conjugate}
The Conjugate Gradient Method (CGM) solves the LSP with a complexity of $O(N)$ and is the fastest known classical solver. Therefore, the speed of HHL, which has a complexity of $O(log(N))$, is often compared to the speed of CGM \cite{speedup}. Thus, HHL provides an exponential speedup over the fastest known classical method.

When we solve a system of linear equation, according to (Eq.(\ref{tosolve})),where $A$ is a matrix, $b$ is a vector and $x$ is to be solved. If $A$ is a $2\times2$ matrix and $b$ is $2\times1$ ,then $x$ can be solved easily. But is $A$ is a long matrix, for example $1000,000,000 \times 1000,000,000$ and $b$ is $1000,000,000\times1$ vector, and $N$ in this case is $1000,000,000$. To solve $x$ in Classical Gaussian Elimination method we need $O(N^3)$ speed, where $O$ is omega. In Classical Conjugate Gradient Method  with sparse matrix that contains lots of zeros, it will take $O(N)$ speed. And for HHL Quantum Algorithm, with sparse matrix, it takes $O(log(N))$ speed.

However, according to the paper \cite{ Aaronson2015}, the speed of inner product in HHL Quantum Algorithm is only $log (mn)/\epsilon$ steps when certain amplitudes is obtained and distinguished among other amplitudes, otherwise the speed is only quadratically faster than classical algorithm.

To solve (\ref{tosolve}) in $CGM$ method,
\begin{align}
A
\vec{x}
=
\vec{b}
\tag{\ref{tosolve}}
\end{align}
initial guess of $\vec{x}$ is used as the starting point. The residual is then found and the search direction is determined by finding the steepest descent. This is repeated until a stable condition is met.  

The residual in the $k$\th search is given as,
\begin{align}
R_k
=
b
-
A
\vec{x}_k
\end{align}

The readers do not need to understand CGM to understand HHL. Interested readers may refer to the literature (e.g. \cite{Chandra1975}) for more details.

\EOD

\end{document}